\documentclass{ws-ijmpa}

\newcommand{\ui}{\mathrm{i}}
\newcommand{\ue}{\mathrm{e}}

\newcommand{\UI}{\mathbf{I}}

\newcommand{\ud}{\mathrm{d}}

\newcommand{\re}{\Re}

\newcommand{\vp}{\boldsymbol{\psi}}

\newcommand{\vz}{\mathbf{0}}

\newcommand{\rk}{\mathrm{rank}}

\newcommand{\A}{{\mathbb A}}
\newcommand{\B}{{\mathbb B}}
\newcommand{\D}{{\mathbb D}}

\newcommand{\LM}{{\mathbf{L}}}
\newcommand{\bb}{{\overline{b}}}

\newcommand{\bdm}{\begin{displaymath}}
\newcommand{\edm}{\end{displaymath}}
\newcommand{\beq}{\begin{equation}}
\newcommand{\eeq}{\end{equation}}
\newcommand{\beqa}{\begin{eqnarray}}
\newcommand{\eeqa}{\end{eqnarray}}
\newcommand{\nn}{\nonumber}
\newcommand{\sep}{\setlength\arraycolsep{2pt}}

\DeclareMathOperator{\csch}{csch}

\def\I{{\rm i}}                  
\def\D{{\rm d}}                  

\newcommand{\deriv}[2]{\frac{\mathrm{d}#1}{\mathrm{d}#2}}

\begin{document}

\markboth{J. M. Harrison and K. Kirsten}
{Vacuum energy of Schr\"odinger operators on metric graphs}

%
\catchline{}{}{}{}{}
%

\title{VACUUM ENERGY OF SCHR\"ODINGER OPERATORS ON METRIC GRAPHS}

\author{JONATHAN M HARRISON}

\address{Department of Mathematics, Baylor University\\ Waco, TX 76798, USA\\
jon\_harrison@baylor.edu}

\author{KLAUS KIRSTEN}

\address{Department of Mathematics, Baylor University\\ Waco, TX 76798, USA\\
klaus\_kirsten@baylor.edu}

\maketitle

\begin{history}
\received{2/12/2011}
\revised{17/1/2012}
\end{history}

\begin{abstract}
We present an integral formulation of the vacuum energy of Schr\"odinger operators on finite metric graphs.
Local vertex matching conditions on the graph are classified according to the general scheme of Kostrykin and Schrader.  While the vacuum energy of the graph can contain finite ambiguities the Casimir force on a bond with compactly supported potential is well defined.
The vacuum energy is determined from the zeta function of the graph Schr\"odinger operator which is derived from an appropriate secular equation via the argument principle.  A quantum graph has an associated probabilistic classical dynamics which is generically both ergodic and mixing.  The results therefore present an analytic formulation of the vacuum energy of this quasi-one-dimensional quantum system which is classically chaotic.

\keywords{quantum graph; vacuum energy; zeta function.}
\end{abstract}

\ccode{PACS numbers: 03.65.Db, 03.70+k, 05.45.Mt, 64.60.aq}

\section{Introduction}	

Quantum graphs have recently become popular as simple models of complex quantum systems as results derived on graphs have a good track record of providing insight in other areas.   Some examples of fields where quantum graphs have been usefully employed include Anderson localization, photonic crystals, microelectronics, the theory of waveguides, quantum chaos, nanotechnology, and superconductivity; see Refs.~\refcite{p:K:GMWPTS} and \refcite{p:K:QG:I&BS} for reviews.  Here we consider a quantum graph model of vacuum energy.  The vacuum energy of a star graph was first described by Fulling, Kaplan and Wilson where they note that for Neummann like matching conditions the Casimir effect is repulsive.\cite{p:FKW:VERCFQSG}  Subsequently the vacuum energy of the Laplacian on a general metric graph was formulated as a periodic orbit sum\cite{p:BHW:MAVEQG} and in terms of the vertex matching conditions by the authors.\cite{p:HK:VE,p:HK:ZFQG}  One interesting feature of quantum graphs is that they possess a probabilistic classical dynamics that is chaotic.  Such a mathematical model of vacuum energy for a spectrum of a chaotic system follows in the spirit of the Riemannium model\cite{p:LMB:R} where the height up the critical line of the zeros of the Riemann zeta function provides the spectrum.  Both the height of the zeros of the Riemann zeta function on the critical line and the  spectrum of a generic quantum graph (with time reversal symmetry broken by a magnetic field) are distributed like the eigenvalues of a large random Hermitian matrix from the Gaussian Unitary Ensemble,\footnote{When the spectra are unfolded so they have mean spacing one.} a typical feature of the spectrum of a system that is classically chaotic.\cite{p:KS:POTQG}

In order to regularize the vacuum energy we employ the \emph{spectral zeta function},
\begin{equation}\label{eq:spec zeta}
    \zeta(s,\gamma)={\sum_{j=0}^{\infty}} (\gamma + E_j)^{-s} \ ,
\end{equation}
where $\gamma\in \mathbb{R}$ is some spectral parameter and the sum runs over the spectrum of the Schr\"odinger operator $0< E_0 \leqslant E_1 \leqslant \dots $; the condition that the spectrum be positive can be relaxed by introducing some  additional technicalities.\cite{p:KM:FDGSLP}
 The zeta function is constructed from a secular equation whose positive roots $k_j$ satisfy $E_j=k_j^2$.\footnote{The degeneracy of the roots and energy eigenvalues also coincide.}    From the secular equation the zeta function is derived via the argument principle, as in Ref. \refcite{p:HKT:SD&ZF} where the authors apply the zeta function to derive the spectral determinant of the Schr\"odinger operator on a graph with general vertex matching conditions.

The vacuum energy is defined as
\begin{equation}\label{eq:E_c}
    \mathcal{E}_c=\lim_{\epsilon \to 0} \frac{\mu^{2\epsilon}}{2} \zeta ( \epsilon-1/2,0 ) \ ,
\end{equation}
which is formally $\frac{1}{2} \sum_{j=0}^{\infty} \sqrt{E_j}$.  $\mu$ is an arbitrary parameter with the units of mass, which is included as $\zeta(s,\gamma)$ will generically have a pole at $s=-1/2$.  Taking the limit $\epsilon \to 0$,
\begin{equation}\label{eq:E_c expansion}
    \mathcal{E}_c= \frac{1}{2} \mathrm{FP} \, \zeta(-1/2,0) +\frac{1}{2} \left( \frac{1}{\epsilon} +\ln \mu^2 \right) \mathrm{Res} \, \zeta(-1/2,0) +O(\epsilon) \ ,
\end{equation}
where $\mathrm{FP}$ and $\mathrm{Res}$ denote the finite part and residue respectively.
From the zeta function we obtain the following general result for the Casimir force on a bond $\beta$ of the graph.
\begin{theorem}\label{thm:F_c}
For the Schr\"odinger operator on a graph with local vertex matching conditions defined by a pair of matrices $\A$ and $\B$, with $\A\B^\dagger=\B\A^\dagger$ and $\rk (\A,\B)=2B$, and potential functions $V_b(x_b)\in C^\infty$ for $b=1,\dots, B$ the Casimir force on a bond $\beta$ where
$\mathrm{supp} \, V_\beta \subset (0,L_\beta)$ is given by
\begin{equation}
    \mathcal{F}_{c}^\beta := -\frac{\partial}{\partial L_\beta} \mathcal{E}_c
    = \mathcal{F}^\beta_{c, \textrm{Dir}}-\frac 1 {2\pi} \int_0^\infty \frac{\partial}{\partial L_\beta} \log F(\ui t)\, \ud t \ ,
\end{equation}
where $F(\ui t)= \det \left( \A + \B M(\ui t) \right)$ and $\mathcal{F}^\beta_{c, \textrm{Dir}}$ is the Casimir force on the interval $[0,L_\beta]$ with Dirichlet boundary conditions.
\end{theorem}
%
$M({\ui t})$ is a $2B \times 2B$ matrix defined in terms of the solution to a boundary value problem on the intervals $[0,L_b]$; see Section \ref{sec:secular}.
The secular equation whose solutions $k$ are square roots of the eigenvalues of the Schr\"odinger operator reads $F(k)=0$.  In the following we assume $F(0)\neq 0$, a condition that can be relaxed.\cite{p:HKT:SD&ZF}
The Casimir force on an interval with Dirichlet boundary conditions is defined in Eq.
(\ref{eq:F_c Dir compact}).


The article is organized as follows.
Section \ref{sec:graph} introduces the Sch\"odinger operator on a graph.  Section \ref{sec:secular} describes the secular equation we use for the quantum graph. In Section \ref{sec:zeta} we recall our recent results\cite{p:HKT:SD&ZF} for the zeta function of the graph Schr\"odinger operator.  In Section \ref{sec:Casimir}
we evaluate the vacuum energy of the graph and prove Theorem \ref{thm:F_c}.  

\section{Graph model}\label{sec:graph}

\begin{figure}[!ht]
  \begin{center}
  \includegraphics[width=5cm]{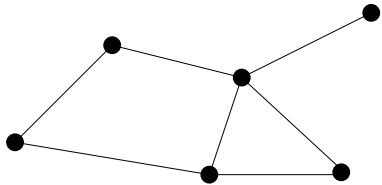}
  \caption{\it A finite graph with $6$ vertices and $7$ bonds.\label{fig:graph}}
  \end{center}
\end{figure}

A graph $G$ is a collection of \emph{vertices} $v=1,\dots,V$ and \emph{bonds} $b=1,\dots,B$,
see Fig. \ref{fig:graph}.
Each bond connects a pair of vertices $b=(v,w)$ and we use $o(b)=v$ to denote the origin vertex of $b$ and $t(b)=w$ for the terminal vertex.  $\bb = (w,v)$ will denote the reversed bond with the origin and terminus exchanged.  We consider undirected graphs with $B$ undirected bonds, $b$ and $\bb$ will denote to the same physical bond with the label $b$ referring to the bond with $o(b)<t(b)$.
$m_v$ denotes the degree or valency of $v$, the number of bonds meeting at $v$.
To determine the valency $b$ and $\overline{b}$ are a single bond.

In a metric graph each bond $b$ corresponds to an interval $[0,L_b]$.  For the coordinate $x_b\in [0,L_b]$,  $x_b=0$ at $o(b)$ and $x_b=L_b$ at $t(b)$, $L_b$ is the \emph{length} of bond $b$.
We also use the coordinate $x_{\bb}=L_b-x_b$ which measures the distance to $x_b$ from the terminal vertex of $b$.
A function $\psi$ on $G$ consists of a collection of functions $\{ \psi_b(x_b) \}_{b=1,\dots,B}$ on the set intervals.
The choice of coordinate on a physical bond enforces the relation $\psi_b(x_b)=\psi_\bb(x_\bb)=\psi_\bb(L_b-x_b)$.
The Hilbert space of $G$ is then
\begin{equation}\label{eq:Hilbert space}
    {\mathcal H} = \bigoplus_{b=1}^B L^2 \bigl( [0,L_b] \bigr) \ .
\end{equation}

The Schr\"odinger equation on the interval $[0,L_b]$ reads,
\begin{equation}\label{eq:eigenproblem}
    \left( \ui \frac{\ud}{\ud x_b} +A_b \right)^2 \psi_b(x_b) + V_b(x_b) \psi_b(x_b) = k^2 \psi_b(x_b) \ .
\end{equation}
$A_b$ is a vector potential on $b$ and for consistency $A_\bb=-A_b$ as the direction is reversed when changing coordinate.\footnote{A vector potential $A_b(x_b)$ that depends on $x_b$ can be made constant by a gauge transformation.}
%
Matching conditions at the set of vertices are specified by a pair of $2B\times 2B$ matrices $\A$ and $\B$ according to
\begin{equation}\label{eq:matching conditions}
    \A \vp + \B \hat{\vp}=\vz \ ,
\end{equation}
where
\begin{eqnarray}\label{eq:vertex values}
    \vp&=&\big(\psi_1(0),\dots,\psi_B(0),\psi_1(L_1),\dots,\psi_B(L_B)\big)^T \ , \\
    \hat{\vp}&=&\big(D_{1} \psi_1(0),\dots, D_{B} \psi_B(0), D_{\overline{1}} \psi_1(L_1),\dots,D_{\overline{B}} \psi_B(L_B)\big)^T \ .
\end{eqnarray}
Here, $D_b := \frac{\ud}{\ud x_b} -\ui A_b$ is the covariant derivative, so $\hat{\vp}$ is the vector of inward pointing covariant derivatives at the ends of the intervals.
The Schr\"odinger operator is self-adjoint if and only if $(\A,\B)$ has maximal rank and $\A\B^\dagger=\B\A^\dagger$.\cite{p:KS:KRQW}  Such pairs of matrices classify the self-adjoint realizations of the Schr\"odinger operator.  (See Ref. \refcite{p:K:QG:I} for an alternative unique classification scheme for general vertex matching conditions.)  

For example, if we specify that the wavefunction is continuous at a vertex, $\delta$-type matching conditions, a choice of matrices $\A$ and $\B$ that encodes for this is
\begin{equation}\label{eq:A B delta}
  \mathbb{A}_\delta  =
  \left(
    \begin{array}{ccccc}
      -\lambda & 0 & 0 & \cdots & 0\\
      -1 & 1 & 0 & \cdots & 0\\
      0 & -1 & 1 & \cdots & 0\\
      \vdots & \vdots &  & & \vdots \\
      0 & 0 & 0 & \cdots  & 1
    \end{array}
  \right)
  \hspace{0.5cm} \mbox{and} \hspace{0.5cm}
  \mathbb{B}_\delta  =
  \left(
    \begin{array}{ccccc}
     1 & 1 & \cdots & 1 & 1\\
     0 & 0 & \cdots & 0 & 0 \\
     0 & 0 & \cdots & \cdots & 0 \\
     \vdots & \vdots &  & & \vdots \\
     0 & 0 & \cdots & \cdots & 0
    \end{array}
  \right)
\end{equation}
 where $\lambda$ fixes the strength of the coupling at the vertex.  If $\lambda=0$ the incoming derivatives at the vertex sum to zero, Neumann like matching conditions.


We further restrict the possible matching conditions to \emph{local matching conditions} where the matrices $\A$ and $\B$ only relate values of functions and their derivatives on the intervals where they meet at a vertex and where the matrices are independent of the metric structure of the graph, namely the bond lengths and the spectral parameter $k$.  Local matching conditions will be required subsequently to use the argument principle.

\section{Secular equation}\label{sec:secular}

We first consider the simplest graph, an interval $[0,L]$ with Dirichlet boundary conditions, $\psi(0)=\psi(L)=0$.
Let $f(x;-k^2)$ be a solution of the Schr\"odinger Eq. (\ref{eq:eigenproblem}) such that $f(0;-k^2)=1$ and $f(L;-k^2)=0$.  A second linearly independent solution is denoted
$\bar f(\bar x;-k^2)$ where $\bar x=L-x$ and $\bar f$ satisfies
$\bar f(0;-k^2)=1$ and $\bar f(L;-k^2)=0$.  An alternative set of solutions are $u_{k}(x;-k^2)$ and $\bar u(\bar x;-k^2)$ satisfying
$u(0;-k^2)=0$ and $u'(0;-k^2)=1$.
The function $\bar u$ is then,
\begin{equation}
    \bar u(\bar x;-k^2) = { - \frac{ f(x;-k^2) }{ f'(L;-k^2) } }
   \:.
\end{equation}
$k>0$ is a square root of the eigenvalues of the Schr\"odinger operator iff
\begin{equation}
  \label{eq:Dirischlet secular}
   \bar u(L;-k^2) = -1/ f'(L;-k^2) = 0
  \: ,
\end{equation}
which defines a secular equation for the Dirichlet problem on the interval.  For a graph with Dirichlet boundary conditions at the end of each interval $[0,L_b]$ the set of intervals are effectively detached and the spectrum is the union of the spectrum of each of the intervals.
Hence a secular equation for the graph with Dirichlet boundary conditions is
\begin{equation}
  \label{eq:Dirichlet graph secular}
   \prod_{b=1}^B \big( -1/ f_b'(L_b;-k^2) \big) = 0
  \: .
\end{equation}

If we now consider the Schr\"odinger equation on a general graph with matching conditions defined by the matrices $\A$ and $\B$ the component of a wavefunction with energy $k^2$ on bond $b$ can be written
\begin{equation}\label{eq:wavefn on b}
    \psi_{b} (x_b,-k^2) = c_b \, f_{b} (x_b;-k^2) \, \ue^{\ui A_b x_b}  + c_\bb \, f_{\bb} (x_\bb;-k^2)
    \, \ue^{\ui A_\bb  x_\bb} \ ,
\end{equation}
where $f_b$ is a solution of the eigenvalue problem on the interval $[0,L_b]$ as defined previously.
The vectors of values of the wavefunction and its inward covariant derivatives are then
\begin{equation}\label{eq:vertex values 2}
    \vp=(c_1,\dots,c_B,c_{\overline{1}},\dots, c_{\overline{B}} )^T   \quad \textrm{and} \quad
    \hat{\vp}=M(-k^2)\vp \ ,
\end{equation}
where $M(-k^2)$ is the $2B\times 2B$ matrix
\begin{equation}\label{eq:defn of M}
    M_{ab}=\delta_{a,b} \, f'_{b}(0;-k^2) - \delta_{a,\bb} \, f'_{\bb } (L_b;-k^2) \ue^{\ui A_b L_b} \ .
\end{equation}
Substituting in the vertex matching conditions we find that $k^2$ is an eigenvalue of the Schr\"odinger operator iff
$k$ is a solution of the secular equation
\begin{equation}\label{eq:secular}
    F(k):= \det \left( \A +\B M(-k^2) \right) =0 \ .
\end{equation}
Notice that the function $F(z)$, where $z=k+\ui t$, has poles along the real axis at points where $1/ f'_{\bb } (L_b;-z^2)$ vanishes, i.e. at the eigenvalues of the graph with Dirichlet boundary conditions at the ends of the intervals.

\section{Zeta function}\label{sec:zeta}

\begin{figure}[!ht]
  \begin{center}
  \setlength{\unitlength}{1cm}
    \begin{picture}(13,3.5)
    \put(0,0){\includegraphics[scale=0.75]{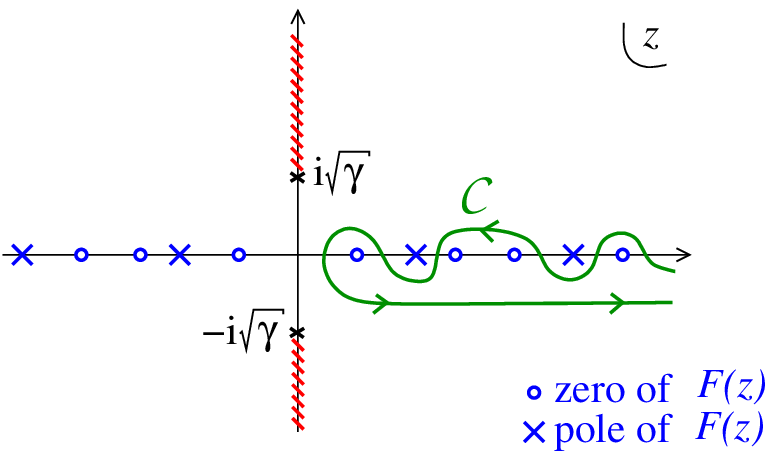}}
    \put(6.5,0){\includegraphics[scale=0.75]{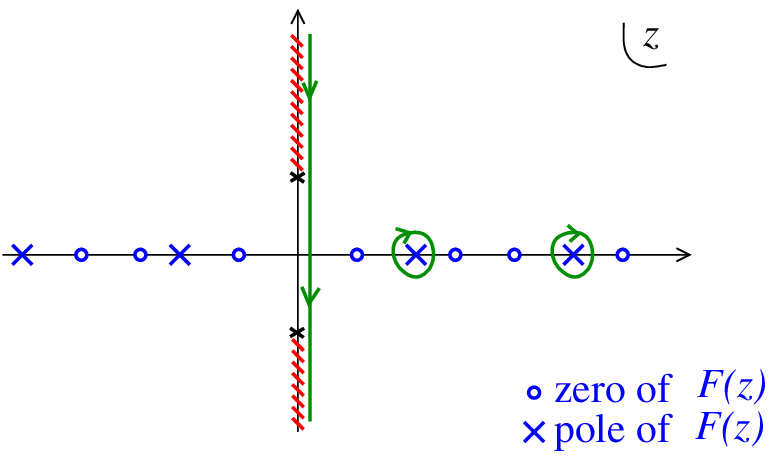}}
    \put(0,3){a)}
    \put(6.5,3){b)}
    \end{picture}
  \caption{\it The contours used to evaluate the graph zeta function,
    (a) before, and (b) after, the contour deformation.
    The two branch cuts are represented by hashed lines.
  \label{fig:contours}}
  \end{center}
\end{figure}

We construct the zeta function of the graph from the secular equation using the argument principle.\cite{p:HKT:SD&ZF}
For an interval with Dirichlet boundary conditions the secular equation reads $u(L;-k^2) = 0$.
The function $u(L;-k^2)$ has no poles and we can represent the zeta function with a contour integral where the contour $\mathcal{C}$ encloses the positive real axis, namely
\begin{equation}
  \zeta_\mathrm{Dir}(s,\gamma)
  = \frac{1}{2\I\pi} \int_\mathcal{C}
  (z^2+\gamma)^{-s}\deriv{}{z}\log\left( u(L;-z^2) \right) \, \D z \ .
\end{equation}
Transforming the contour $\mathcal{C}$ to the imaginary axis $z=\ui t$, Figure \ref{fig:contours}(b), we obtain
\begin{equation}
   \zeta_\mathrm{Dir} (s,\gamma)  = \frac{\sin\pi s}{\pi}
  \int_{\sqrt{\gamma}}^\infty
  (t^2-\gamma)^{-s}\,
  \deriv{}{t}\log\left( u(L;t^2) \right)\, \D t
  \:,
\end{equation}
a representation which is valid in the strip $1/2<\re s<1$; see Ref. \refcite{p:HKT:SD&ZF} for details.
For example, for a free particle where $V(x)=0$, $u(x;-k^2)=\sin (kx)/k$ so $u(L;t^2)=\sinh (tL)/t$.

For a graph with Dirichlet boundary conditions at the ends of all the intervals the zeta function is the sum of the Dirichlet zeta functions on the individual bonds.  A graph with general local vertex matching conditions has secular equation $F(z)=0$ defined in Eq.
(\ref{eq:secular}).
$F$ has zeros on the positive real axis at square roots of eigenvalues of the operator and poles at eigenvalues of the graph with Dirichlet conditions.  We assume we are in the generic case where these are distinct.  The contour $\mathcal{C}$ is then chosen so that it encloses the zeros of $F$ and excludes the poles on the positive real axis, see Fig. \ref{fig:contours}(a).  Transforming the contour to the imaginary axis, Fig. \ref{fig:contours}(b), we see that the zeta function has two components,
\begin{equation}\label{eq:Im plus poles}
    \zeta(s,\gamma)=\zeta_\mathrm{Im}(s,\gamma) +\sum_{b=1}^B \zeta^b_{\mathrm{Dir}}(s,\gamma) \ .
\end{equation}
$\zeta^b_{\mathrm{Dir}}$ is the zeta function of the graph with Dirichlet boundary conditions at the ends of the intervals which comes from the poles of $F$ and $\zeta_\mathrm{Im}$ is the contribution of the integral along the imaginary axis,
\begin{equation} \label{eq:zeta Im}
    \zeta_\mathrm{Im}(s,\gamma)
    = \frac{\sin \pi s}{\pi} \int_{\sqrt{\gamma}}^{\infty} (t^2-\gamma)^{-s} \frac{\ud}{\ud t} \log F(\ui t) \,  \ud t \ ,
\end{equation}
which converges in the strip $1/2<\re s<1$.

In order to evaluate the vacuum energy we need to extend the range of validity to include $s=-1/2$.  The restriction to $\re s > 1/2$ comes from the asymptotic behavior of $F$ as $t\to \infty$.  The asymptotics of $f'_{b}$ can be analyzed using the WKB method,\cite{p:HKT:SD&ZF} and we find
\begin{equation}\label{eq:g'(L) asymp early}
    f'_{b}(0;t^2) \underset{t\to\infty}{\sim}  -t +\sum_{j=1}^\infty s_{b,j}(0)t^{-j} \ ,
\end{equation}
and $f'_b(L_b;t^2)$ vanishes exponentially.  The functions $s_{b,j}(x_b)$ are determined by the recurrence relation
\begin{equation}\label{eq:recursion}
  s_{b,j+1} (x_b)=  \frac{1}{2} \left( s_{b,j}'(x_b) +\sum_{i=0}^j s_{b,i}(x_b) s_{b,j-i}(x_b) \right) \ ,
\end{equation}
where $s_{b,-1}(x_b) = - 1$, $s_{b,0}(x_b) = 0$ and  $s_{b,1}(x_b) = - V_b(x_b)/2$.
Using (\ref{eq:g'(L) asymp early}) we can write the expansion of $F(\ui t)$ up to exponentially suppressed terms as
\begin{equation}\label{eq:f expansion}
    F(\ui t) \underset{t\to\infty}{\sim} \det \Big(\A+\B(-t\UI+D(t))\Big) \ ,
\end{equation}
where
\begin{equation}\label{eq:D(t)}
  D(t)= \sum_{j=1}^\infty t^{-j}\,  \textrm{diag} \left\{   s_{1,j}(0),\dots, s_{B,j}(0), s_{\overline{1},j}(0),\dots, s_{\overline{B},j}(0) \right\} \ .
\end{equation}
This expansion of $F(\ui t)$ fits with the free particle case investigated previously in Refs. \refcite{p:HK:VE} and \refcite{p:HK:ZFQG} where, in the absence of a potential, $D(t)=\vz$ and $F(\ui t) \sim \det (\A-t\B)$.
In general the expansion takes the form
\begin{equation}\label{eq:generic F expansion}
    F(\ui t) \underset{t\to\infty}{\sim} \sum_{j=0}^\infty c_j \,  t^{2B-j} \ .
\end{equation}
Let us denote by $c_{N}$ the first nonzero coefficient in the expansion, so $N=0$ if $c_0=\det \B \neq 0$, and by $c_{N+J}$ the second non-zero coefficient.  Hence,
\begin{equation}\label{eq:log F expansion}
    \log  F(\ui t) \underset{t\to\infty}{\sim} \log (c_N t^{2B-N}) +\frac{c_{N+J}}{c_N t^J} +O(t^{-(J+1)}) \ .
\end{equation}
Adding and subtracting these first two terms in the expansion of $F$ we obtain a representation of $\zeta_{\mathrm{Im}}(s,\gamma)$ convergent in the strip $-(J+1)/2 < \re s<1$, namely{\sep
\begin{eqnarray}\label{eq:zeta Im extended}
\zeta_{\textrm{Im}}(s,\gamma)&=&\frac{\sin \pi s}{\pi} \int_{\sqrt{\gamma}}^\infty (t^2-\gamma)^{-s} \frac{\ud}{\ud t}
    \left[ \log \Big( F(\ui t)t^{N-2B} / c_N \Big) -\frac{c_{N+J}}{c_N\, t^J} \right] \, \ud t \nn \\
    &&  +
\frac{(2B-N)\sin \pi s}{2\pi s}\gamma^{-s} -\frac{J c_{N+J}\, \Gamma(s+J/2)}{2c_N \, \Gamma(s)\Gamma(J/2+1)}\gamma^{-(2s+J)} \ .
\end{eqnarray}}
%

We must also represent the Dirichlet zeta function of each interval in a strip containing $s=-1/2$.
The large $t$ asymptotics of $\log u_b(L_b;t^2)$ are given by\cite{p:FKM:PMP}
\begin{equation}
  \log   u_b(L_b;t^2) \underset{t\to\infty}{\sim}
  t L_b - \log (2t) + d_b \, t^{-1} +O(t^{-2})
  \ ,
\end{equation}
where $d_b=\frac{1}{2} \int_0^{L_b} V(x_b) \, \ud x_b$.  Subtracting and adding the first three terms we obtain the representation{\sep
\begin{eqnarray}\label{eq:zeta Dir extended}
  \zeta^b_{\mathrm{Dir}}(s,\gamma)  &=& \frac{\sin\pi s}{\pi}
  \int_{\sqrt{\gamma}}^\infty
  (t^2-\gamma)^{-s}\,
  \deriv{}{t}
  \left[
     \log\left( u_b(L_b;t^2) \right) - t L_b + \log(2t) -d_b \, t^{-1}
  \right]\, \D t \nn \\
  && + L_b \frac{\Gamma(s-1/2)}{2\sqrt\pi \Gamma(s)}\gamma^{\frac12-s}
  -\frac12\gamma^{-s} - d_b \frac{2\Gamma(s+1/2)}{\sqrt\pi \Gamma(s)} \gamma^{-\frac{1}{2}-s}
  \ ,
\end{eqnarray}
}valid in the strip $-1<\re s<1$.


\section{Casimir effect}\label{sec:Casimir}

Following Eq. (\ref{eq:E_c expansion}), to evaluate the vacuum energy we identify{\sep
\begin{eqnarray}\label{eq:FP zeta Im}
    \mathrm{FP} \, \zeta_{\textrm{Im}} (-1/2,0) &=&
    -\frac{1}{\pi} \int_{1}^\infty t \frac{\ud}{\ud t}
    \left[ \log \Big( F(\ui t)t^{N-2B} / c_N \Big) -\frac{c_{N+J}}{c_N\, t^J} \right] \, \ud t
    \nn \\
    && -\frac{1}{\pi} \int_{0}^1 t \frac{\ud}{\ud t}
    \left[ \log \Big( F(\ui t)\Big) \right] \, \ud t \ .
\end{eqnarray}
}In the generic case when $J=1$ there is a pole of  $\zeta(s,0)$ at $s=-1/2$ and
\begin{equation}\label{eq:Res zeta Im}
    \mathrm{Res} \, \zeta_{\textrm{Im}} (-1/2,0) = \frac{c_{N+1}}{2\pi c_N} \ .
\end{equation}
For the interval $[0,L_b]$ with Dirichlet boundary conditions{\sep
\begin{eqnarray}\label{eq:FP zeta Dir}
    \mathrm{FP} \, \zeta^b_{\textrm{Dir}} (-1/2,0) &=&
    -\frac{1}{\pi} \int_{1}^\infty t \frac{\ud}{\ud t}
    \left[ \log\left( u_b(L_b;t^2) \right) - t L_b + \log(2t) -d_b \, t^{-1} \right]\,
    \D t
    \nn \\
    &&
    -\frac{1}{\pi} \int_{0}^1 t \frac{\ud}{\ud t}
    \left[ \log\left( u_b(L_b;t^2) \right) \right]\,
    \D t \ ,
\end{eqnarray}
}and $\mathrm{Res} \, \zeta^b_{\textrm{Dir}} (-1/2,0)= d_b/\pi$.  Combining these results and using Eqs.
(\ref{eq:E_c expansion}) and (\ref{eq:Im plus poles}) we have obtained a representation of the vacuum energy of the graph Schr\"odinger operator.

If we assume that the potential on a bond is compactly supported and consider the Casimir force on that bond we can obtain a more explicit description of the Casimir effect on the graph.  For a bond $\beta$ let
\begin{equation}\label{eq:compact sup}
    \mathrm{supp}\, V_\beta  \subset (0,L_\beta) \ ,
\end{equation}
then the Casimir force on the bond $\beta$ is finite and can be evaluated without ambiguity as the asymptotic behavior of $F(\ui t)$ is independent of $L_\beta$.
We define the Casimir force on bond $\beta$ for the compactly supported potential as $F_{c}^{\beta} =-\frac{\partial}{\partial L_\beta} \mathcal{E}_c$, then
\begin{equation}\label{eq:F_c compact}
    \mathcal{F}_{c}^\beta
    = \mathcal{F}^\beta_{c, \textrm{Dir}}-\frac 1 {2\pi} \int_0^\infty \frac{\partial}{\partial L_\beta} \log F(\ui t)\, \ud t \ .
\end{equation}
$\mathcal{F}^\beta_{c, \textrm{Dir}}$ is the force on the interval $[0,L_\beta]$ with Dirichlet boundary conditions,
\begin{equation}\label{eq:F_c Dir compact}
    \mathcal{F}_{c,\textrm{Dir}}^\beta
    =- \frac  1 {2\pi} \int_0^\infty \frac{\partial}{\partial L_\beta} \log \left( \frac{u_\beta(L_\beta;t^2)}{ \ue^{L_\beta t}} \right)\, \ud t \ .
\end{equation}
Hence we obtain Theorem \ref{thm:F_c}.


For example, with no potential on the graph
\begin{equation}\label{eq:F_c Dir compact}
    \mathcal{F}_{c,\textrm{Dir}}^\beta
    =- \frac  1 {2\pi} \int_0^\infty \frac{\partial}{\partial L_\beta} \log \left( \frac{\sinh (L_\beta t)}{t \ue^{L_\beta t}} \right)\, \ud t
    =-\frac{\pi}{24 L^2} \ ,
\end{equation}
   a result that can also be obtained directly from the spectrum $E_j=\left( \frac{\pi j}{L} \right)^2$ of the Laplace operator on an interval with Dirichlet boundary conditions, where $\zeta_\textrm{Dir}(s,0)=\left(\frac{\pi}{L}\right)^{-2s} \zeta_\textrm{R}(2s)$.

\begin{figure}[!htb]
  \begin{center}
  \includegraphics[width=2cm]{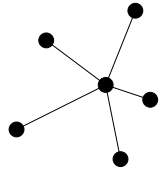}
  \caption{\it A star graph with $5$ bonds.\label{fig:star}}
  \end{center}
\end{figure}

  For a free particle we can also write $f_{b}$ explicitly,
\begin{equation}\label{eq: V=0 f}
    f_{b}(x_b;-k^2)= \frac{\sin k(L_b-x_b)}{\sin k L_b} \ , \quad \textrm{ hence } \quad
    f_{b}'(x_b;t^2)= \frac{-t\cosh t(L_b-x_b)}{\sinh t L_b} \ .
\end{equation}
The matrix $M$ then has the form
\begin{equation}\label{V=0 M}
    M(t^2)= t \left( \begin{array}{cc}
    -\coth t \LM & \csch t \LM \\
     \csch t \LM & - \coth t \LM \\
    \end{array} \right) \ ,
\end{equation}
where $\coth t \LM =\textrm{diag} \{ \coth t L_1 , \dots, \coth t L_B \}$ and $\csch t \LM$ is defined similarly.  If we choose a particular graph like the star graph, Fig. \ref{fig:star}, with Dirichlet boundary conditions at the vertices of degree one and $\delta$-type coupling at the central vertex, such matching conditions can be encoded in the $2B\times 2B$ matrices,
\begin{equation}\label{eq: A B star}
    \A=\left( \begin{array}{cc}
    \UI_B &0 \\
    0& \A_\delta\\
    \end{array} \right) \qquad
    \B=\left( \begin{array}{cc}
    0 &0 \\
    0& \B_\delta\\
    \end{array} \right)
\end{equation}
with $\A_\delta$ and $\B_\delta$ defined as in Eq. (\ref{eq:A B delta}).
%
%
Substituting in (\ref{eq:secular}) and evaluating the determinant we find
\begin{equation}\label{eq:F star}
    F(\ui t)=- \lambda -\sum_{b=1}^B \coth tL_b \ ,
\end{equation}
and the Casimir force on a bond $\beta$ is given by,
\begin{equation}\label{eq:F_c V=0}
    \mathcal{F}_{c}^\beta
    = -\frac{\pi}{24 L^2}-\frac 1 {2\pi} \int_0^\infty \frac{\partial}{\partial L_\beta} \log F(\ui t)\, \ud t \ .
\end{equation}

If the condition that the potential $V_\beta$ is compactly supported is relaxed, finite renormalization ambiguities in the Casimir force remain.
These are the standard ambiguities as they occur in self-interacting $(\lambda \Phi^4)$-theories and they can be dealt with in a routine way.\cite{zuber}


%
%

\section*{Acknowledgments}

The authors would like to thank Guglielmo Fucci and Christophe Texier for stimulating discussions.
KK is supported by National Science Foundation grant PHY--0757791.  JMH was supported by the Baylor University summer sabbatical program.


\end{document}